\documentclass{aa}  

\usepackage{graphicx}
\usepackage{siunitx}
\usepackage{color}
\usepackage{txfonts}
\usepackage[normalem]{ulem}
\usepackage[modulo]{lineno}
\begin{document}
\title{Gaia view of primitive inner-belt asteroid families}
\subtitle{Searching for the origins of asteroids Bennu and Ryugu}

   \author{M. Delbo\inst{1}
    \and
         C. Avdellidou\inst{1}
          \and
          K. J. Walsh\inst{2}%\fnmsep\thanks{Just to show the usage of the elements in the author field}
        }

   \institute{Université C\^ote d’Azur, CNRS–Lagrange, Observatoire de la C\^ote d’Azur, CS 34229 – F 06304 NICE Cedex 4, France\\
              \email{delbo@oca.eu}
         \and
             Southwest Research Institute, 1050 Walnut St. Suite 300, Boulder, CO, 80302, USA\\
             %\email{c.ptolemy@hipparch.uheaven.space}
            % \thanks{The university of heaven temporarily does not accept e-mails}
             }

   \date{Received September 15, 1996; accepted March 16, 1997}

% \abstract{}{}{}{}{}
% 5 {} token are mandatory
 
  \abstract
  {}
   {Near-Earth asteroids Ryugu and Bennu, were visited, characterised, and sampled by the Hayabusa2 and OSIRIS-REx missions, where remote sensing data and sample return analysis showed that both asteroids have primitive, hydrated and organic-rich compositions. The dark families of the inner main belt that belong to the spectroscopic C-complex have been claimed to be the sources of both Ryugu and Bennu, hence there have been large efforts to spectroscopically characterise them by ground-based observations.}
   {Here we used the Gaia Data Release 3 (Gaia DR3) asteroid reflectance spectra in order to characterise the 11 known inner main belt C-complex families (Chaldaea, Chimaera, Clarissa, Erigone, Eulalia, Klio, Polana, Primordial, Sulamitis, Svea, Tamara), using space-borne visible-light spectroscopic observations. 
For each family we extracted the family members that have known geometric visible albedo values and Gaia DR3 data, and we created an average reflectance spectrum per family, between 370 and 950~\SI{}{\nano\meter}. 
 These averages were then compared with the ground-based visible spectroscopic surveys of the same families, and to Bennu's and Ryugu's space and ground-based spectra in the same wavelength range.}
   {Gaia DR3 reflectance spectra of the dark asteroid families of the inner main belt are in general consistent with previous findings. The only exception is the case of the Svea family: previous surveys classified its members as B-types, whereas the average reflectance spectrum from Gaia DR3 is similar to a C-type. We also showed that the Polana and the Eulalia families can be distinguished in the wavelength region 370 -- 500~\SI{}{\nano\meter}.
   %
   %In the same blue region, 
    Among all the primitive inner main belt families, we found that the average reflectance spectra of the Eulalia and Polana families are the most similar to those of Bennu and Ryugu, respectively. In particular, Eulalia family's average spectrum is a good match to Bennu's in the wavelength range 450 -- 800~\SI{}{\nano\meter}, while beyond 800~\SI{}{\nano\meter} the spectrum of Bennu is bluer than that of Eulalia. Moreover, the spectrum of the Polana family has the smallest discrepancy (smallest $\chi^2$) against the spectrum of Ryugu, although this match is formally unsatisfactory (reduced $\chi^2 \sim$ 1.9).}
  {}

   \keywords{minor planets, asteroids:general --
            astronomical databases:miscellaneous --
                techniques: spectroscopic}

   \maketitle
%
%________________________________________________________________
%\linenumbers
\section{Introduction}
The main belt consists of over one million known asteroids. These have a plethora of compositions and values of albedo, ranging from few to several tens of percent. The two largest compositional classes are those broadly identified by the so-called spectroscopic S- and C-complexes \citep{demeo2015}. These spectral complexes are also separated in albedo, with the S- and C-complex asteroids having, in general, geometric visible albedo ($p_V$) values smaller than or larger than 0.12, respectively \citep{delbo2017,Ferrone2023}.

Recent theories of asteroid formation invoke an \textit{in situ} (or quasi-\textit{in situ}) origin of the S-complex bodies \citep{walsh2011,Raymond_2017SciA}. On the other hand, C-complex asteroids are supposed  to be exogenous to the main belt. It is generally thought that these bodies accreted from the dust of the protoplanetary disk in the giant planet region \citep{Warren_2011,kruijer2017,Nakamura_2022}, were later transported, and finally implanted into the main belt from larger heliocentric distances. A handful of mechanisms have been proposed for this process, including the rapid growth of the giant planets scattering nearby planetesimals \citep{Raymond_2017Icar}, the migration of the giant planets amidst the solar nebula \citep{walsh2011,Pirani2019} and likely some contributions from the giant planet instability \citep{Levison_2009,DavidV_2016AJ}.

%EDITBYKJW%One of the possible mechanism responsible for this transport and implantation is the inward migration of Jupiter down to 1.5~au, and its subsequent outward migration \citep{walsh2011} during the very early phases of the solar system evolution (e.g. $\lesssim$5~Myr). 

After the original asteroids \citep[also known as planetesimals;][]{delbo2017} were accreted (for the S-complex) or implanted (for the C-complex) in the main belt, collisions with other asteroids throughout the history of the solar system have broken several of them, which in turn created families of asteroid fragments \citep{nesvorny2015}. Hence, the large majority of asteroids are collisional fragments of the original ones \citep{delbo2017,dermott2018,Ferrone2023}. 

This large population of small fragments can reach the near-Earth space via well-established dynamical routes and eventually arrive to Earth as meteorites \citep[][and references therein]{greenwood2020}. The carbonaceous chondrite meteorites (CCs) are typically associated with the C-complex asteroids. The study of CCs has shown that the C-complex asteroids carry a primitive composition, enriched in water and organics, and were possibly responsible for delivering the ingredients of life to primeval Earth via impacts \citep{chyba1992}. On the other hand, despite the fact that the C-complex asteroids are a majority in the main belt, the CCs are a small fraction in our meteoritic inventory \citep{norton2008}. This is reasonable if we consider the filtering by Earth's atmosphere. Due to the fact that C-complex asteroid materials are mechanically quite weak and some of them with large porosity \citep{grott2019,ballouz2020,cambioni2021}, it is very difficult to survive the atmospheric passage at the typical velocities that meteoroids hit our planet. 

The above have also been confirmed by Hayabusa2 (JAXA) and OSIRIS-REx (NASA) sample return missions, which have visited, characterised, and sampled the near-Earth C-complex asteroids (162173) Ryugu \citep{kitazato2019} and (101955) Bennu \citep{lauretta2019}, respectively. Recent sample analysis has shown that Ryugu is an aqueous altered asteroid, with a CI (Ivuna-like) \citep{yokoyama2023}, organic-rich composition \citep{ito2022,naraoka2023,yabuta2023}. Moreover, Ryugu samples provide an excellent example of primitive solar system material that is "clean" from terrestrial contamination. Although Bennu's sample has not arrived on Earth yet, OSIRIS-REx spectroscopic data from the visible to thermal infrared wavelengths have shown that the surface is consistent with an aqueously altered CM chondrite \citep{hamilton2019,Kaplan2020Sci,Simon_2020Sci}. 

It is established that, based on the orbits of Ryugu and Bennu, both asteroids escaped the inner-main belt (IMB) via the $\nu_6$ secular resonance \citep{campins2010b,campins2013,bottke2015} and reached the near-Earth space about 5 \citep{okazaki2023} and 1.75~Myr~ago \citep{ballouz2020}, respectively. Therefore, their source regions (i.e., C-complex asteroid families), should be located in the inner main belt and could possibly have different compositions.

In the inner main belt 11 dark C-complex asteroid families of different sizes and ages are known to exist \citep{walsh2013,nesvorny2015,delbo2017}, namely: Polana, Eulalia, Erigone, Chaldaea, Chimaera, Clarissa, Klio, Sulamitis,  Svea, Tamara, and the so-called Primordial one \citep[the age of a family is given by the epoch of the collision that created it;][]{nesvorny2015}. Since 2016, the majority of these families have been the subject of dedicated ground-based visible and near-infrared spectroscopic observations to characterise their composition, report on the level of their homogeneity, and compare to each other. This effort has been mostly driven by the need to constrain the source regions of Ryugu and Bennu, to give astronomical context for the returned samples \citep[e.g.,][]{lauretta2015}.

Specifically, it has been reported that any slope difference between the Polana and Eulalia family members in the visible light range (VIS) is within 1$\sigma$ uncertainty \citep{deleon2016}, while they are neither distinguishable in the near-infrared range (NIR) \citep{pinilla_alonso2016}, nor in the near-ultraviolet (NUV) \citep{Tatsumi2022A&A...664A.107T}. Visible spectroscopic observations of Erigone, Clarissa, Sulamitis, Klio, Chaldaea, and Svea \citep{morate2018,morate2019}, showed that the slopes of Clarissa and Svea family members agree with the Polana family members, and, similarly, Sulamitis family members' slopes to Erigone ones. Despite their difference in VIS slopes, NIR observations showed that Sulamitis and Klio members are close to Polanas \citep{arredondo2020,arredondo2021b}.
In addition, Klio and Chaldaea show a complementarity in VIS slopes that together could match the Erigone's spectral slope distribution, while in the NIR appear to have extremely similar slopes. It is proposed that these findings could indicate a common origin to these families \citep{arredondo2021a}, an idea that can be supported by the fact that both families overlap in the space of proper orbital semi-major axis vs. inverse diameter ($a$ vs. 1/$D$). Another interesting result is that the dark IMB families have been divided into two broad groups, the so-called "blue" families with blue to moderate slopes and no sign of hydration, and the "red" families, which include objects with the 700~\SI{}{\nano\meter} absorption band present \citep{morate2019}.

In this work, our goal is to study the IMB dark primitive families by exploiting hundreds of VIS spectra from Gaia Data Release 3 (DR3) catalogue \citep{galluccio2022}, in order to understand their differences and similarities. In this way we will obtain a view of their original asteroids (planetesimals). In section 2 we will present the datasets that include the family membership and the Gaia DR3 spectra of these populations;  in sections 3 our analysis and results, while in section 4 the discussion on our findings, including a  comparison with the ground-based spectroscopic studies.

%__________________________________________________________________

\section{Data}

\subsection{Gaia DR3 spectroscopic data}
We made use of the reflectance spectra that were derived from asteroid spectroscopic observations obtained by Gaia between 5 August 2014 and 28 May 2017 and were published in June 2022 as part of the DR3 \citep{galluccio2022}. This dataset consists of mean reflectance spectra in the VIS wavelength range of 60,518 Solar System objects (SSOs). The reflectance spectra were acquired by two low-resolution slit-less spectrophotometers on board Gaia, the blue and red spectrophotometers (BP and RP), which are respectively optimised for the blue and red part of the spectrum. Specifically, the BP spans the wavelength range from 330 to 680~\SI{}{\nano\meter} and the RP covers the range from 640 to 1050~\SI{}{\nano\meter}. The spectral resolution of each spectrophotometer is a function of wavelength, and varies from 4 to 32~\SI{}{\nano\meter} pixel$^{-1}$ for the BP and 7 to 15~\SI{}{\nano\meter} pixel$^{-1}$ for the RP \citep{galluccio2022,carrasco2021,jordi2010}. When an asteroid transited on the focal plane of Gaia at a given epoch, each spectrophotometer measured photons at every wavelength to create `epoch spectra'. For each asteroid, each epoch spectrum was divided by the mean spectrum of a series of trusted solar analogue stars \citep[see Table~C.1 of][]{galluccio2022} in order to create "epoch reflectances". Given that the wavelength range of both instruments overlaps in the 650-680~\SI{}{\nano\meter} interval, the two epoch reflectances were merged to create a full epoch reflectance. To each asteroid was finally associated a unique mean reflectance spectrum obtained by averaging several epoch reflectances, spanning the visible wavelength range from 374 to 1034~\SI{}{\nano\meter} in 16 discrete wavelength bands \citep{galluccio2022}. A `reflectance\_spectrum\_flag' (hereafter RSF) number was also associated with each band, assessing the estimated quality of the band. In some cases, the merging of the epoch spectra taken by each spectrophotometer was not perfect, which can lead to the creation of artefact bands \citep{galluccio2022}. Thus, caution must be taken when analysing the mean reflectance spectra in the overlapping wavelength interval. In a similar way, the bluest and reddest data bands of Gaia spectra could also be affected by systematics due to the low efficiency of the spectrophotometers in these bands. They were not always flagged but they need to be taken with caution as well \citep{galluccio2022,galinier2023}.

\subsection{C-complex inner main belt families}
We retrieved the asteroid membership for seven out of 11 IMB C-complex families from \cite{nesvorny2015} (see Table~\ref{table:1}). The latter family identification within the main belt asteroid population is, by construction, conservative, meaning that good separation between the families is ensured, limiting therefore the number of family interlopers. This is very good for our purposes of studying similarities and differences between families in the same region of the main belt.  For the Polana and Eulalia families we used the membership as it was defined in \cite{walsh2013}. For the Tamara family, which is located in the high-inclination Phocaea region, we used the membership of \cite{novakovic2017}. Finally, for the so-called Primordial family we used the membership defined by \cite{delbo2017}. We retrieved the $p_V$-values, where applicable, from the Minor Planet Physical Properties Catalogue (MP3C)\footnote{https://mp3c.oca.eu}. Specifically, for each asteroid the  $p_V$-values reported by the MP3C are averaged values of the published geometric visible albedo determination (a weighting factor equal to the inverse of square of the published formal uncertainty of each albedo determination is used by the MP3C for their averaging). 

%__________________________________________________________________
\section{Analysis \& Results}

\subsection{Removal of family interlopers}
It is important to remember that the identification of family members based on clustering in orbital elements \citep{nesvorny2015} or correlations between the orbital proper semi-major axes and the $1/D$ \citep[V-shape criterion][]{bolin2017,delbo2017} is a rough expression of the true membership \citep{nesvorny2015,Ferrone2023}: in fact, asteroids unrelated to a family that happen to have values of proper elements and $1/D$-values within the orbital and/or $1/D$ range of that family are grouped together with the true members. These objects identified as "false positive" are typically called family interlopers \citep{nesvorny2015}. Interlopers can be distinguished amongst family members by their anomalous spectral properties or albedo values compared to the majority of the family members.

To filter out potential family interlopers and ensure the cleanest sample for our spectral analysis of each family, we applied a series of criteria on the values of the $p_V$, spectral slope ($s$), and reflectance difference $R_z-R_i$ of family members. 
First of all, we required the family members to have $p_V$-values reported in the literature, and be $p_V$ < 0.12. In addition, we required the Gaia DR3 spectra to have -5 < $s$ < 6~\%/100~\SI{}{\nano\meter}, and -0.2 < $R_z-R_i$ < 0.185. The selected $p_V$ threshold has been shown to separate the dark from the bright asteroids respectively belonging to the C- and S-complexes \citep{delbo2017}. In particular, 88\% of the C-complex asteroids are contained within the asteroid population with $p_V$<0.12 \citep{delbo2017}. Moreover, the ranges of spectral parameters mentioned above define the boundaries of the C- and B-classes and therefore of the spectroscopic C-complex \citep{demeo2013}. The application of these criteria results in reducing the number of members of each family included in our analysis; the latter number is reported, for each family, in Table~\ref{table:1}. 

The spectral parameters were calculated following the method of \cite{galluccio2022}, in two steps. First, the spectral slope was determined as the angular coefficient of the best-fit straight line to the reflectance data between 450 and 600~\SI{}{\nano\meter}. Then, the reflectance difference $R_z-R_i$ was obtained by fitting a natural smoothing spline, $S(\lambda)$, (Python3 module CSAPS; smoothing coefficient of $5 \times 10^{-7}$) and then by assuming that $R_z-R_I = S(\lambda_z) - S(\lambda_i)$, where $\lambda_z$ = 893.2~\SI{}{\nano\meter} and $\lambda_i$ =748.0~\SI{}{\nano\meter}.

\begin{table*}
\caption{Input dataset.}             
\label{table:1}     
\centering                          
\begin{tabular}{l |l l l c c c c c c |l}        
\hline\hline                 
Family & $N_{members}$ & $N_{DR3}$ & $N_{filtered}$ & $p_V$ &	$\sigma$$_{p_V}$ &	$s$ & $\sigma$$_{s}$ & $R_z-R_i$ & $\sigma$$_{R_z-R_i}$ & Ref.\\  
\hline\hline
Chaldaea &		132	&  40   &	32	&	0.067	&	0.019	&+1.30	&	2.53	&	0.066	&	0.035	& 	\cite{nesvorny2015}\\
Chimaera &	108	&  18   & 11 	&	0.054	&	0.014	&+2.65	&	1.39	&	0.099	&	0.041	&	\cite{nesvorny2015}\\     
Clarissa &		179	&  2	   & 2 	&	0.069	&	0.002	&-1.00	&	-	&	0.036 	&	-		&	\cite{nesvorny2015}\\
Erigone & 		1776	&  142 & 82 	&	0.056 	&	0.017	&+1.75	&	2.14	&	0.053	&	0.035	&	\cite{nesvorny2015}\\
Eulalia & 		1624	&   248& 205 	& 	0.057	&	0.012	&+1.02	&	2.03	&	0.038	&	0.042	&	\cite{walsh2013}\\
Klio & 		330	& 88    & 72 	&	0.067 	&	0.016	&+2.23	&	2.17	&	0.061	&	0.039	&	\cite{nesvorny2015}\\
Polana & 		2037	& 577  & 243 	&	0.058 	&	0.016	&+1.01	&	2.31	&	0.035	&	0.045	&	\cite{walsh2013}\\      
Primordial  & 	118	&  89   & 64 	&	0.061 	&	0.022	&+1.93	&	2.29	&	0.043	&	0.048	&	\cite{delbo2017}\\
Sulamitis & 	303	& 32    & 23 	&	0.056 	&	0.010	&+3.14	&	1.99	&	0.071	&	0.038	&	\cite{nesvorny2015}\\
Svea & 		48	&  4     & 2 	&	0.060 	&	0.019	&+1.42	&	-	&	0.006	&	-		&	\cite{nesvorny2015}\\
Tamara & 		226	&  56   & 56  	&	0.060 	&	0.014	&+1.60	&	2.11	&	0.060  	&	0.042	&	\cite{novakovic2017}\\
\hline                                   
\end{tabular}
\tablefoot{The symbols $N_{members}$, $N_{DR3}$, and $N_{filtered}$ represent, respectively, the number of family members, the number of family members with Gaia DR3 reflectance spectra and the number of family members used in our analysis after the application of the filters described in the main text. The symbols $p_V$, $\sigma$$_{p_V}$, $s$, $\sigma$$_{s}$, $R_z-R_i$, and $\sigma$$_{R_z-R_i}$ represent, respectively, the average geometric visible albedo, the standard deviation of the geometric visible albedo, the reflectance spectral slope between 450 and 750~\SI{}{\nano\meter}, the standard deviation of the slope, the difference between the reflectance at 893.2 and at 748.0~\SI{}{\nano\meter}, and the standard deviation of the difference of reflectance for the distribution of the members of the different families filtered as explained in the text. The column Ref. reports the reference from which we obtained the initial family membership definition.}
\end{table*}

\subsection{Spectra profiles of IMB dark families}
\begin{figure}
   \centering
  \includegraphics[width=.5\textwidth]{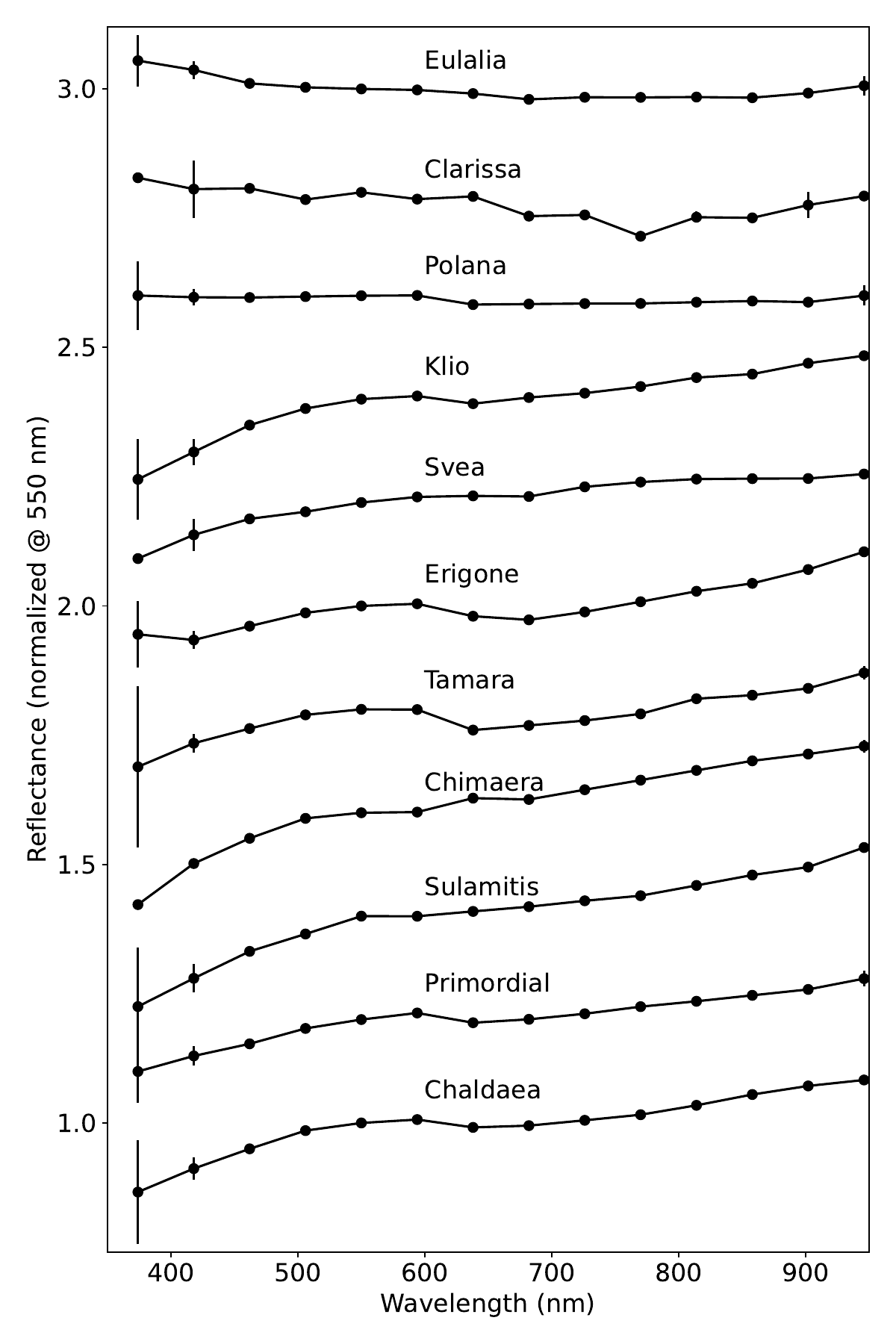}
      \caption{Average Gaia DR3 spectra of the 11 inner main belt dark families that belong to the spectroscopic C-complex.   Spectra are shifted on the reflectance axis for better visibility. The error bars are included, but are smaller than the marker size in most cases.}
         \label{F:allSpec}
   \end{figure}

Once the family members were filtered for potential interlopers, we calculated the average reflectance spectrum of each family.
First, for each reflectance spectrum, we removed those data points with a `reflectance\_spectrum\_flag' (RSF) value $>$ 0, which indicates that those values are suspected to be of poorer quality \citep{galluccio2022}. In addition, we did not take into account all spectra data at 990 and 1034~\SI{}{\nano\meter} because the reflectance at these wavelengths is typically affected by a systematic increase \citep{galluccio2022}.
Next, we calculated the weighted average at each of the 16 wavelength-bands with which the Gaia DR3  reflectance spectra were expressed \citep{galluccio2022}. We also calculated the median average deviation at each wavelength, using $1/\sigma^2$ as weights, where $\sigma$ is the reflectance uncertainty reported in Gaia DR3. After having rejected those data whose distance from the mean is larger than 2.5$\times$ the median average deviation, we recalculated the weighted mean and its uncertainty. The uncertainty of the average reflectance spectra were calculated using a bootstrap technique: namely, we iteratively randomly selected 75\% of the filtered family and recalculated an average reflectance spectrum and, after having reached 1000 iterations, we calculated the standard deviation of the mean spectra at each wavelength. Finally, we applied the correction in the blue region of the spectrum, following the procedure of \cite{tinaut_ruano2023}, where we multiplied by 1.07, 1.05, 1.02, and 1.01 the reflectance values at the wavelengths of 374, 418, 462, and 506~\SI{}{\nano\meter}, respectively. The resulting average reflectance spectra and their uncertainties for each family are shown in Fig.~\ref{F:allSpec}. 

%__________________________________________________________________

\section{Discussion}

There appear to be two classes of dark families in the inner main belt. Specifically, Polana, Eulalia and Clarissa have a blue-to-neutral reflectance spectra, while the other families show a slightly redder reflectance spectra. This pattern has been already noted previously \citep{morate2019}. Within this second group, the Gaia DR3 data of Chaldaea, Primordial, Tamara, Erigone and perhaps Klio families show evidence of the 700~\SI{}{\nano\meter} hydration band. 

In addition, the Gaia DR3 spectra of Polana and Eulalia families can be distinguished in the near-ultraviolet region below 550~\SI{}{\nano\meter}, where Eulalia family appears bluer than Polana (Fig.~\ref{F:polaEulaBennu}). 
This spectral behaviour is also evident specifically on asteroids (142) Polana and (495) Eulalia, the largest members of their respective families (Fig.~\ref{F:polaEulaBennu}). 
A similar behaviour can be found in the ground-based reflectance spectra of asteroids (142) Polana and (495) Eulalia \citep{deleon2016}. In these data, (495) Eulalia turns bluer than (142) Polana at wavelengths shorter than $\sim$450~\SI{}{\nano\meter}.

\subsection{Comparison with ground-based observations}

\begin{figure*}
\centering
\includegraphics[width=1\textwidth]{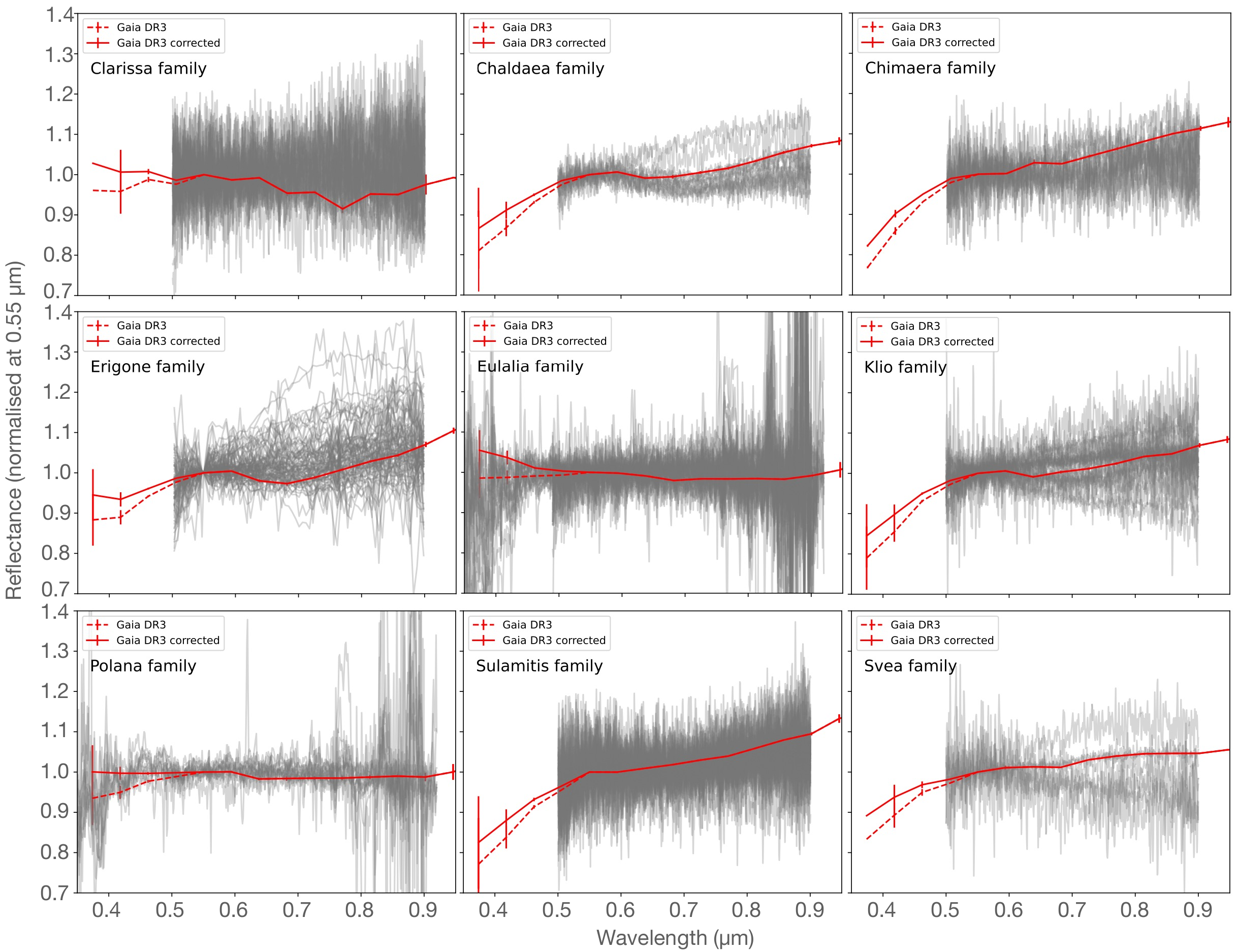}
\caption{Comparison of Gaia DR3 average spectra (red line) for each IMB primitive family to the ground-based visible spectra of the literature (gray lines) from the PRIMASS-L survey (see text). Gaia DR3 reflectance spectra are plotted with and without the correction at the blue most wavelengths of \cite{tinaut_ruano2023}.}
\label{F:PRIMASS-L}
\end{figure*}

In order to compare our results with previous studies, we downloaded the PRIMASS-L Spectra Bundle V1.0 from the NASA PDS archive \citep{pinilla_alonso2021},
%Pinilla-Alonso, N., De Pra, M., de Leon, J., Morate, D., Lorenzi, V., Arredondo, A., Campins, H., Licandro, J., Delbo, M., Cabrera-Lavers, A., Walsh, K., DeMeo, F., Sarid, G. (2021). PRIMASS-L Spectra Bundle V1.0, urn:nasa:pds:gbo.ast.primass-l.spectra::1.0, NASA Planetary Data System, https://doi.org/10.26033/xnfh-np39.
which contains VIS reflectance spectra of several members of the C-complex IMB asteroid families.  In this dataset, asteroids' spectra are already organised by family, with the exception of the Tamara and the Primordial families that are not listed, while the Eulalia and the Polana families are merged together. Hence, we separated Eulalia and Polana families' members on the basis of their position in the $(a, 1/D)$ space, namely, objects with $1/D > -1.7 (a-2.49)$ and $p_V<0.12$ were assigned to the Eulalia family, while objects with $1/D < -1.7 (a-2.49)$ and $p_V<0.12$ were assigned to the Polana family. Finally, for all families, we removed interlopers as reported in their respective publications, as well as the spectra that are classified as D and T types. Figure~\ref{F:PRIMASS-L} shows that the agreement between Gaia DR3 and PRIMASS-L reflectance spectra is in general quite good, with the exception of the Svea family, where the majority of PRIMASS-L spectra have a blue slope, while Gaia DR3 one is redder. One of the possible reason for this discrepancy is that there are only two objects of the Svea family with DR3 reflectance spectra that pass our filters, namely the asteroids (329) and (102626). The former, which is the asteroid Svea itself, has also a red sloped reflectance spectrum in PRIMASS-L, while for the latter there are no literature spectra previous to Gaia DR3. 

We also used a classical $\chi^2$ technique \citep[e.g.,][]{Avdellidou2022} to measure the goodness of fit between the Gaia DR3 and the PRIMASS-L reflectance spectra for the asteroid families of Fig.~\ref{F:PRIMASS-L}. First, for each family, $f$, and for each PRIMASS-L  reflectance spectrum, we determined a fitting natural spline (Python csaps) in order to evaluate PRIMASS-L reflectances at the same wavelengths $\lambda$ of the Gaia DR3 ones. Next, for each family, we calculated a PRIMASS-L mean spectrum $\overline R_f$ and its standard deviation $\overline \sigma_f$. Finally, we calculated a $\chi^2_f = \sum_\lambda \left[ \left (\overline R_f(\lambda) - R_f(\lambda)\right)^2 / \overline \sigma_f^2 \right]$, where $R_f(\lambda)$ is the Gaia DR3 average reflectance spectrum of the family $f$. We also calculated the reduced $\chi^2$ by the classical expression $\tilde \chi^2_f = \chi^2_f / \nu$, where $\nu$ is the number of degrees of freedom \citep{numerical-recipes} that in our case is the number of Gaia DR3 bands used in the expression above minus 1. Results are given in Table~\ref{T:chi2Primass}. Assuming the $\chi^2$ statistic, one can estimate the maximum  $\tilde \chi^2  = 1+\sqrt{2 \nu}/\nu$ \citep{numerical-recipes,Hanus2018Icar..309..297H} within which it is the expected to have 68\% of the cases, i.e., the PRIMASS-L spectra,  assuming that spectral variability is random. Since $1+\sqrt{2 \nu}/\nu$ is equal to 1.5 and 1.4 for the cases with $\nu$ equal to 8 and 12, we can deduce that the Gaia DR3 reflectance spectra fits well those from PRIMASS-L for all the families that the surveys have in common.

\begin{table}
\caption{$\chi^2$-values between Gaia DR3 and PRIMASS-L reflectance spectra for the low-albedo asteroid families of the inner main belt.}             
\label{T:chi2Primass}     
\centering                          
\begin{tabular}{l r c r}        
\hline\hline                 
Family & $\chi^2_f$ & $\nu$ & $\tilde \chi^2_f$\\  
\hline\hline
Chaldaea & 1.69 & 8 & 0.21 \\	
Chimaera & 7.17 & 8 & 0.90\\	
Clarissa & 7.60 & 8 & 0.95 \\	
Erigone & 3.12 & 8 & 0.39 \\	
Eulalia & 5.40 & 12 & 0.45 \\	
Klio &  0.70 &8 & 0.09\\		
Polana & 7.99 & 12 & 0.67 \\	
Primordial  & - & - & - \\
Sulamitis & 1.72 & 8 & 0.22\\	
Svea & 3.23 &8 & 0.41 \\		
Tamara & - & - & - \\	
\hline                                   
\end{tabular}
\tablefoot{There are no spectra for the Primordial and the Tamara families from the PRIMASS-L of \cite{pinilla_alonso2021}.}
\end{table}

Concerning the distinguishability between the Polana and Eulalia families, previous works \citep{deleon2016,pinilla_alonso2016} concluded that members of the two families could not be distinguished on the basis of visible and/or infrared spectroscopy. \cite{Tatsumi2022A&A...664A.107T} also spectroscopically compared the Eulalia and the Polana families (and the Themis family) focusing on the blue (or as they call it the NUV) region (350 -- 550~\SI{}{\nano\meter}) of the spectrum. Namely,  these authors obtained ground-based, low-resolution visible spectra as blue as 350~\SI{}{\nano\meter} of few asteroids belonging to these families. They also used literature spectroscopic data from the eight colour asteroid survey \citep[ECAS,][]{Zellner1985Icar...61..355Z}. \cite{Tatsumi2022A&A...664A.107T} calculated the NUV and VIS spectral slopes by straight line least-square fit to the spectra data of each asteroid between 360 and 550~\SI{}{\nano\meter} and 550 and 850~\SI{}{\nano\meter}, respectively. By comparing the NUV and VIS slopes of the asteroids from the different families, these authors found no significant differences between the Polana and Eulalia families. 

However, using the same data from \cite{Tatsumi2022A&A...664A.107T}, we reached a different conclusion: We used the two-dimensional Kolmogorov–Smirnov test (K–S test) to verify the null hypothesis that the NUV and VIS slopes from the TNG observations reported in Table~4 of \cite{Tatsumi2022A&A...664A.107T} from Eulalia and Polana could come from the same distribution. We found that this null hypothesis could be rejected at 97.2\% confidence level with a K-S test distance of 0.71. Also, we performed the K-S test with a Monte Carlo method between the nominal and random draws from the same data of the asteroids of the Polana family and found that, in 10$^6$ Monte Carlo iterations, the max K-S test distance obtained with random errors is 0.625: i.e., we never obtained 0.71. This means that there is a probability smaller than 10$^{-6}$ to be able to obtain Eulalia NUV and VIS spectral slopes from random errors from the Polana NUV and VIS slopes. If we add the NUV and VIS spectral slopes from Table~4 of \cite{Tatsumi2022A&A...664A.107T} of the ECAS survey, the difference between the asteroids from the Polana and the Eulalia families decreases, but still the two families remain distinguishable. In particular, the K-S test still indicates that the two samples from the Eulalia and from the Polana families are not obtained from the same distribution with a confidence level of 95\% to reject the null hypothesis. We concluded that the data from the work of  \cite{Tatsumi2022A&A...664A.107T} do show that the Polana and Eulalia families can be distinguished, contrary to what was stated by the authors. Independent evidence of the distinction between the colours of the Polana and Eulalia family members is also claimed in a recent work based on the Sloan Digital Sky Survey spectrophometry of asteroids \citep{McClure2022DPS....5421202M}.

\subsection{The source region of the NEAs Bennu and Ryugu}

Previous dynamical and spectroscopic considerations lead to the conclusion that Bennu originated from the low albedo component of what was, once upon a time called the Nysa-Polana clan \citep{campins2010b}. It was later shown that this low albedo component is actually composed by two dynamically distinguishable families named Polana (or New Polana) and Eulalia \citep{walsh2013}. Further investigations assessed that Bennu has about 70 and 30\% probability to originate from the Polana and Eulalia families, respectively \citep{bottke2015}.

\begin{figure}
   \centering
  \includegraphics[width=.5\textwidth]{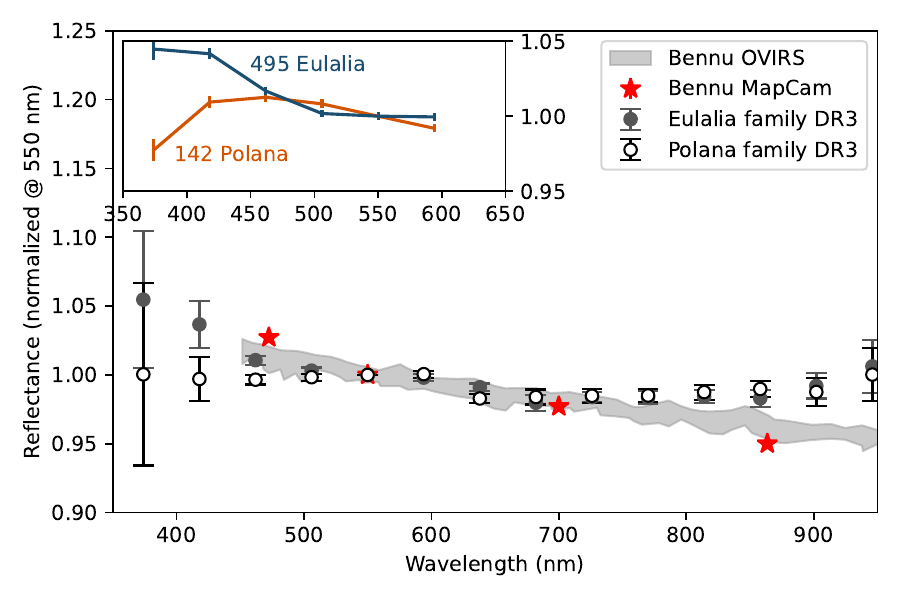}
      \caption{Comparison of the average reflectance spectra of the Eulalia and Polana family from Gaia DR3 and Bennu from \cite{dellagiustina2020}. The subplot also shows the Gaia DR3 spectra of (142) Polana and (495) Eulalia.}
         \label{F:polaEulaBennu}
   \end{figure}

Figure~\ref{F:polaEulaBennu} shows the comparison between the reflectance spectra of Eulalia and  Polana families and the reflectance spectra of Bennu obtained by the OSIRIS-REx Visible and InfraRed Spectrometer (OVIRS) and MapCam instruments on board OSIRIS-REx \citep{dellagiustina2020}.  In the ultraviolet region, it appears that Bennu is marginally more similar to the Eulalia family members than those of the Polana family: namely, Gaia DR3 results slightly favour an origin of Bennu from the Eulalia family because the trend of Bennu's reflectance into the wavelength region bluer than 450~\SI{}{\nano\meter} is more consistent with the reflectance of the Eulalia than of the Polana family (Fig.~\ref{F:polaEulaBennu}). Likewise, Bennu's reflectance trend into the wavelength region bluer than 450~\SI{}{\nano\meter} is more similar to asteroid (495) Eulalia than to the asteroid (142) Polana.

On the other hand, the average reflectance spectra of the Eulalia and Polana families are redder than the reflectance spectrum of Bennu longward $\sim$800~\SI{}{\nano\meter}. It cannot be established with the present Gaia DR3 data whether this is a real effect or an artefact. Indeed, it has been observed \citep{galluccio2022,galinier2023} that Gaia DR3 asteroid spectra tend to have reflectance values higher than the spectra of the same objects collected from ground-based telescopes at wavelengths roughly beyond $\sim$800--900~\SI{}{\nano\meter}.

Figure~\ref{F:polaEulaBennu} subplot also shows that the reflectance spectra of the parent bodies of each respective families are consistent with the families general trend. 
  
We also performed a $\chi^2$ analysis between the reflectance spectra of Bennu, obtained by the OSIRIS-REx mission (Fig.~\ref{F:polaEulaBennu}), and the average reflectance of the families from the Gaia DR3. Namely, we constructed a mean reflectance spectrum $R_B(\lambda)$ and its standard deviation from smoothing spline representations  (Python csasp) of the reflectance spectra from OVIRS and MapCam \citep{dellagiustina2020}. Next, we calculated the reduced $\chi^2$, $\tilde \chi^2_{B,f} = (1/\nu) \sum_\lambda \left[ \left (R_B(\lambda) - R_f(\lambda)\right)^2 / \overline \sigma_f^2 \right]$, where $R_f(\lambda)$ is the Gaia DR3 average reflectance spectrum of the family $f$, as before, and $\nu$ is the degrees of freedom, i.e., the number of wavelength datapoints - 1. 
The lower the value of $\tilde \chi^2_{B,f} $, the better the spectroscopic match is.  Firstly, we considered all data points in the wavelength range between 450 and 950~\SI{}{\nano\meter}, as shown in Fig.~\ref{F:polaEulaBennu} (OVIRS and MapCam spectra do not cover wavelengths bluer than 450~\SI{}{\nano\meter}). We found $\tilde \chi^2_{B,f}$-values of  3.5 and 6.0 for the Eulalia and the Polana families, respectively; all other families producing   $\tilde \chi^2_{B,f}$-values $>$ 76. This result further indicates that the Eulalia is the first and Polana is the second family more spectroscopic alike Bennu.  Given that the spectroscopic match between Eulalia (and Polana) and Bennu, is visually better in the bluer region of the spectrum (Fig.~\ref{F:polaEulaBennu}), we also calculated $\tilde \chi^2_{B,f}$  in different wavelength ranges obtained from the aforementioned one by reducing the upper bound from 950 to 800~\SI{}{\nano\meter}. We found that the values of  $\tilde \chi^2_{B,f}$ decrease monotonically as the upper bound decreases, obtaining values below 1 for the Eulalia family when the upper wavelength bound is reduced to 800~\SI{}{\nano\meter}. The corresponding values are reported in Table~\ref{T:chi2Bennu}. We also found that Polana is the family always providing the second best match. It is common to accept reduced $\chi^2$ values $< (1 + \sqrt{2 \nu} / \nu)$ \citep{numerical-recipes}: only the Eulalia family satisfies this constraint.  
%\sout{The values of $\tilde \chi^2_{B,f} $ are reported in Table~\ref{T:chi2BennuRyugu}. None of the values are below or around 1, which indicates that the spectral matching is never ideal. In this case, it is appropriate to modify the 1$\sigma$ bound rule to require that $\tilde \chi^2_{B,f}  < \tilde \chi^2_{B,f MIN} \left ( 1 + \sqrt{2 \nu}/\nu \right )$, where in our case $\sqrt{2 \nu}/\nu$ is equal to 0.45 for all families.  Table~\ref{T:chi2BennuRyugu} shows that the Eulalia family has an average Gaia reflectance spectrum with the smallest spectral discrepancy with respect to that of Bennu. The spectral discrepancy between Bennu and the Polana family is still marginally acceptable (0.97$\sigma$), while all other families are outside the 1$\sigma$ bound of 3.9. }

\begin{table}
\caption{Reduced $\chi^2$-values between Gaia DR3 average reflectance spectra of the low-albedo asteroid families of the inner main belt and the average reflectance spectrum of Bennu. The $\chi^2$-values are calculated here on the wavelength range between 450 and 800~\SI{}{\nano\meter}.  The table is sorted by the $\chi^2$-value, from the smallest to the largest.}             
\label{T:chi2Bennu}     
\centering                          
\begin{tabular}{l | r c }
\hline            
Family & $\tilde \chi^2_{B,f}$ & $\nu$ \\
  \hline\hline
Eulalia         &     0.99  &  6 \\ %    1.58    2.15    2.73
Polana          &     3.34  &  6 \\ %    1.58    2.15    2.73
Clarissa        &    10.00  &  6 \\ %    1.58    2.15    2.73
Tamara          &    25.33  &  6 \\ %    1.58    2.15    2.73
Erigone         &    68.13  &  6 \\ %    1.58    2.15    2.73
Chaldaea        &    88.03  &  6 \\ %    1.58    2.15    2.73
Klio            &   105.47  &  6 \\ %    1.58    2.15    2.73
Sulamitis       &   124.81  &  6 \\ %    1.58    2.15    2.73
Chimaera        &   752.72  &  6 \\ %    1.58    2.15    2.73
Primordial      &   791.36  &  6 \\ %    1.58    2.15    2.73
Svea            &  1540.66  &  6 \\ %    1.58    2.15    2.73
\hline
\end{tabular}
\end{table}

\begin{table}
  \caption{Reduced $\chi^2$-values between Gaia DR3 average reflectance spectra of the low-albedo asteroid families of the inner main belt and the average reflectance spectra of Ryugu. The $\chi^2$-values are calculated here on the wavelength range between 370 and 950~\SI{}{\nano\meter}.  The table is sorted by the $\chi^2$-value, from the smallest to the largest. }            
\label{T:chi2Ryugu}     
\centering                          
\begin{tabular}{l | r c }
\hline            
Family & $\tilde \chi^2_{R,f}$ & $\nu$ \\
\hline\hline                 
Polana          &     1.91  & 12 \\ %    2.32    2.73    3.13
Clarissa        &     3.30  & 12 \\ %    2.32    2.73    3.13
Eulalia         &     3.37  & 12 \\ %    2.32    2.73    3.13
Tamara          &     3.79  & 12 \\ %    2.32    2.73    3.13
Klio            &     5.72  & 12 \\ %    2.32    2.73    3.13
Erigone         &     5.75  & 12 \\ %    2.32    2.73    3.13
Chaldaea        &     6.72  & 12 \\ %    2.32    2.73    3.13
Sulamitis       &    12.17  & 12 \\ %    2.32    2.73    3.13
Primordial      &    25.33  & 12 \\ %    2.32    2.73    3.13
Chimaera        &   117.90  & 12 \\ %    2.32    2.73    3.13
Svea            &   606.61  & 12 \\ %    2.32    2.73    3.13
\hline
\end{tabular}
\end{table}

\begin{figure}
   \centering
  \includegraphics[width=.5\textwidth]{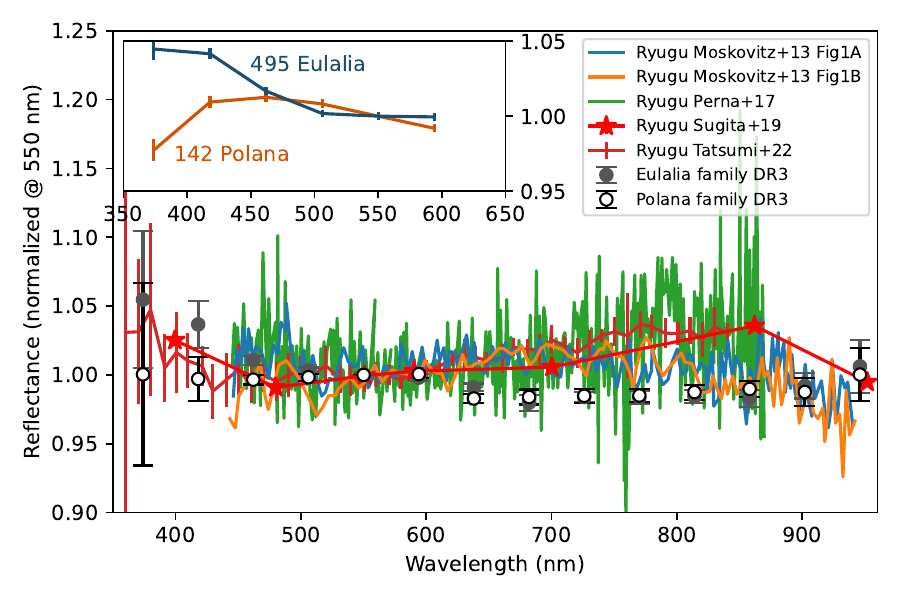}
  \caption{Comparison of the average reflectance spectra of the Eulalia and Polana family from Gaia DR3 and Ryugu from \cite{sugita2019}, \cite{moskovitz2013}, and \cite{perna2017}. The subplot also shows the Gaia DR3 spectra of (142) Polana and (495) Eulalia.}
         \label{F:polaEulaRyugu}
\end{figure}

Figure~\ref{F:polaEulaRyugu} shows the comparison between the reflectance spectra of Eulalia and Polana families and the reflectance spectra of Ryugu obtained by the Optical Navigation Camera (ONC) instrument on board the Hayabusa2 \citep{sugita2019}, along with ground-based spectra \citep{moskovitz2013,perna2017}. In the blue and ultraviolet regions, Ryugu shows a reflectance spectra that are, in principle, compatible with both Polana and Eulalia families average spectra.
On the other hand, at wavelengths longer than 600~\SI{}{\nano\meter}, the average Gaia DR3 reflectance spectra of Polana and Eulalia families have somewhat lower reflectance values compared to the Ryugu spectra measured from Hayabusa2. However, they appear still marginally within the error-bars of Ryugu's ground-based spectra. 

As for Bennu, we performed a $\chi^2$ analysis between the mean reflectance spectra of Ryugu, $R_R(\lambda)$, calculated from averaging smoothing spline fits to the spectra obtained by the Hayabusa2 mission and ground based telescopes (Fig.~\ref{F:polaEulaRyugu}), and the average reflectance spectra of each family, $f$, from the Gaia DR3, $R_f(\lambda)$: namely,  $\tilde \chi^2_{R,f}~=~(1/\nu) \sum_\lambda \left[ \left (R_R(\lambda) - R_f(\lambda)\right)^2 / \overline \sigma_f^2 \right]$.
Considering all the data points in the wavelength range between 370 and 950~\SI{}{\nano\meter}, we obtained  the  $\tilde \chi^2_{R,f} $-values reported in Table~\ref{T:chi2Ryugu} . The family with the spectrum producing the minimum $\chi^2$ value is Polana, followed by the Clarissa and the Eulalia families.  However, this reduced $\chi^2$ value is never close to 1 (the lowest value is $\sim$1.9) for the different spectral ranges here considered, indicating that this spectral match is never formally satisfactory. Contrary to the case of Bennu, calculating the $\tilde \chi^2_{R,f}$ in shorter wavelength intervals, such as between 450 and 800~\SI{}{\nano\meter}, does not lower the minimum $\tilde \chi^2_{R,f} $-value.

%\sout{As in the case of Bennu, also for Ryugu, we cannot find an ideal spectral match because $\tilde \chi^2_{R,f} > 1$. The minimum $\tilde \chi^2_{R,f}$-value corresponds to the Polana family, which is the one offering the least discrepant matching. The Eulalia family average reflectance spectrum still gives a marginally acceptable match (0.9$\sigma$) with all the other families having reduced $\chi^2$ very much outside the 1$\sigma$ bound of 4.42.}

%__________________________________________________________________
\section{Conclusions}
Using the unprecedented Gaia DR3 sample of visible asteroid spectra, we created the average reflectance spectra of the 11 dark primitive asteroid families of the IMB that belong to the spectroscopic C-complex.

In this work, we reported on their similarities, but also their differences, and we compared nine of them with earlier ground-based spectroscopic results. Gaia DR3 provides invaluable reflectance information between 370 and 500~\SI{}{\nano\meter}, a region that is poorly studied so far from the ground, but can indicate differences among the C-complex population. 

Consistently with previous results, we found that there appear to be two spectroscopic groups within these families: Eulalia, Polana and Clarissa are spectroscopically bluer than all the others. Within the redder groups of families, Erigone, Tamara, Chaldaea and the Primordial, appear to have a 700~\SI{}{\nano\meter} absorption band, indicative of hydration.  

We found that the Eulalia and Polana adjacent families can be spectroscopically distinguished in the wavelength region between 370 to 500~\SI{}{\nano\meter}, where Eulalia is bluer than Polana.

%The primitive dark inner main belt families are of great scientific importance as they are the source regions of Bennu and Ryugu, the two visited and sampled near-Earth asteroids, which are found hydrated and organic abundance.
The comparison of the average family spectra with those of the NEAs Bennu and Ryugu showed that these asteroids are spectroscopically more similar to Eulalia and Polana, respectively, than to other families. 

In particular, we found that the Gaia DR3 average spectrum of the Eulalia family is a good spectroscopic match for Bennu's spectrum in the wavelength range between 450 and 800~\SI{}{\nano\meter}. On the other hand, Polana is the family which average DR3 spectrum has the smallest reduced $\chi^2$ when compared to the average spectrum of Ryugu.  

%__________________________________________________________________

\begin{acknowledgements}
We acknowledge support from the ANR ORIGINS (ANR-18-CE31-0014). 
This work has made use of data from the European Space Agency (ESA) mission Gaia (\url{https://www.cosmos.esa.int/gaia}), processed by the Gaia Data Processing and Analysis Consortium (DPAC, \url{https://www.cosmos.esa.int/web/gaia/dpac/consortium}). Funding for the DPAC has been provided by national institutions, in particular the institutions participating in the Gaia Multilateral Agreement.
This work is based on data provided by the Minor Planet Physical Properties Catalogue (MP3C) of the Observatoire de la C\^ote d'Azur.
\end{acknowledgements}

\bibliographystyle{aa}
\bibliography{references.bib,KJW.bib}

\end{document}